\documentclass[10pt]{article}
\usepackage{amsfonts}
\def\Journal#1#2#3#4{{#1} {\bf #2}, #3 (#4)}
\newcommand{\beqn}{\begin{equation}}
\newcommand{\eeqn}{\end{equation}}
\newcommand{\ba}{\hspace*{-5pt}\begin{array}}
\newcommand{\ea}{\end{array}}
\newcommand{\p}{\partial}
\newcommand{\ord}{{\rm ord}\;}
\newcommand{\ds}{\displaystyle}
\newtheorem{theo}{Theorem}
\newtheorem{cor}{Corollary}
\begin{document}

\begin{center}
{\large {\bf On time-dependent symmetries and formal symmetries of
evolution equations}}\\[5mm]
{\sc Artur Sergyeyev}\\
Institute of Mathematics of NAS of Ukraine, \\
3 Tereshchenkivs'ka Str., 252004 Kyiv, Ukraine\\
E-mail: arthurser@imath.kiev.ua,
arthur@apmat.freenet.kiev.ua
\end{center}

\begin{abstract}
We present the explicit formulae, describing the
structure of symmetries and formal symmetries of any scalar
(1+1)-dimensional evolution equation. Using these results, the
formulae for the leading terms of commutators of two symmetries
and two formal symmetries are found. The generalization of these
results to the case of system of evolution equations is also
discussed.
\end{abstract}

\section{Introduction}
It is well known that provided scalar (1+1)-dimensional evolution
equation possesses the infinite-dimensional commutative Lie
algebra of {\it time-independent} generalized (Lie -- B\"acklund)
symmetries, it is either linearizable or integrable via inverse
scattering transform (see e.g.
\cite{sok}, \cite{s}, \cite{o} for the survey of known
results and \cite{mik} for the generalization to (2+1)
dimensions). The existence of such algebra is usually proved by
exhibiting the recursion operator \cite{o} or
mastersymmetry \cite{fu}. But in order to possess the latter, the
equation in question must have some (possibly nonlocal) {\it
time-dependent} symmetries. This fact is one of the main reasons
of growing interest to the study of whole algebra of {\it
time-dependent} symmetries of evolution
equations \cite{fu1}, \cite{mbcf}.

However, to the best of author's knowledge, there exist almost no
results, describing the structure of this algebra for the generic
scalar evolution equation (even in 1+1 dimensions) without any
initial conjectures about the specific properties of this
equation, like the existence of formal symmetry, Lax pair,
mastersymmetry, etc.\footnote{Nevertheless, let us mention
 the papers of Flach \cite{flach} and Magadeev \cite{magad}.}
Although Vinogradov {\it et al.} \cite{v} had
outlined the general scheme of study of local and nonlocal
symmetries of evolution equations, it remained unrealized in its
general form until now.

In this paper we put the part of this scheme into life and go even
further. Namely, in Theorems 1 and 2 we describe the general
structure of local symmetries and formal symmetries of scalar
evolution equation, and in Theorems 3 and 5 we present the
formulae for the leading terms of the commutators of two
symmetries (of sufficiently high order) and of two formal
symmetries. We present also the generalization of these results to
the case of systems of evolution equations.

\section{Basic definitions and known facts}
We consider the scalar $1+1$-dimensional evolution
equation
\begin{equation} \label{eveq}
 \partial u / \partial t =F(x, t, u, u_{1}, \dots, u _{n}),
 \quad n \geq 2,
\quad \p F/\p u_{n} \neq 0,
\end{equation}
where $u_l=\partial ^{l} u /\partial x^{l}$, $l=0,1,2, \dots$,
$u_{0} \equiv u$, and the symmetries of this equation, i.e. the
right hand sides $G$ of evolution equations
\begin{equation} \label{sym}
 \partial u / \partial \tau =G(x,t,u, u_{1}, \dots, u _{k}),
\end{equation}
compatible with equation~(\ref{eveq}). The greatest number $k$
such that $\p G/\p u_{k} \neq 0$ is called the order of symmetry
and is denoted as $k=\ord G$. If $G$ is independent of $u,u_{1},
\dots$, we assume that $\ord G=0$. Let $S^{(k)}$ be the space of
symmetries of order not higher than $k$ of (\ref{eveq}) and $S =
\bigcup_{j=0}^{\infty} S^{(k)}$. $S$ is Lie algebra with respect
to the so-called Lie bracket \cite{sok}, \cite{o}
\[
\{ h,r \} = r_{*} (h) - h_{*} (r),
\]
where for any sufficiently smooth function $f$ of $x, t, u$,
$u_1, \dots, u_s$ we have introduced the notation
\[
f_{*} = \sum\limits_{i=0}^{s} \p h /\p u_{i} D^{i}, \:
\nabla_{f} =\sum\limits_{i=0}^{\infty} D^{i} (f) \p/\p u_{i},
\]
where $\ds D= \p /\p x + \sum\limits_{i=0}^{\infty} u_{i+1} \p/\p u_{i}$.

$G$ is symmetry of Eq.(\ref{eveq}) if and only if  \cite{o}
\beqn \label{comp}
\p G /\p t = - \{ F, G \}.
\eeqn

In many examples Eq.(\ref{eveq}) is quasilinear, so let
\beqn
\label{ql} n_{0} = \left\{
\begin{array}{l}
\max(1-j,0), \: \mbox{if} \:\: \p F/\p u_{n-i} =\phi_{i} (x,t),
\quad i=0,\dots,j,\\ 2 \:\; \mbox{otherwise.}
\end{array}
\right.
\eeqn

It is known  \cite{o} that for any $G \in S$, $\ord G = k \geq
n_{0}$, we have
 \beqn \label{lead}
  \p G/\p u_{k} = c_{k}(t) \Phi^{k/n},
 \eeqn
where $c_{k}(t)$ is a function of $t$ and $\Phi = \p F/\p u_{n}$.

It is also well known  \cite{sok} that for any sufficiently smooth
functions $P,Q$ of $x,t,u,u_{1}, u_{2}, \dots$ the relation $R
=\{P, Q\}$ implies
\beqn \label{lin}
 R_{*} = \nabla_{P} (Q_{*}) - \nabla_{Q}(P_{*}) +
[Q_{*}, P_{*}],
\eeqn
 where $\ds \nabla_{P}(Q_{*}) \equiv \sum\limits_{i,j=0}^{\infty} D^{j}(P)
{\ds \frac{\p^{2} Q}{\p u_{j} \p u_{i}}} D^{i}$ and likewise for
$\nabla_{Q} (P_{*})$; $[\cdot,\cdot]$ stands for the usual
commutator of linear differential operators.

In particular, Eq.(\ref{comp}) yields
\beqn \label{cr3}
 \p G_{*} /\p t \equiv (\p G/\p t)_{*} = \nabla_{G} (F_{*}) -
\nabla_{F}(G_{*}) + [F_{*}, G_{*}].
 \eeqn

Equating the coefficients at $D^{s}$, $s=0,1,2, \dots$ on right
and left hand sides of Eq.(\ref{cr3}), we obtain
\beqn \label{cr4}
 \ba{l}
  {\ds \frac{\p^{2} G}{\p u_l \p t}} =
\sum\limits_{m=0}^{n} D^{m} (G) {\ds \frac{\p^{2} F}{\p u_{m} \p
u_{l}}} - \sum\limits_{r=0}^{k} D^{r} (F) {\ds \frac{\p^{2} G}{\p
u_{r} \p u_{l}}} \\[4mm] \quad + \sum\limits_{j=\max(0,
l+1-n)}^{k} \sum\limits_{i=\max(l+1-j,0)}^{n} \Bigl[ {\rm
C}_{i}^{i+j-l} {\ds \frac{\p F}{\p u_{i}}} D^{i+j-l} \Bigl( {\ds
\frac{\p G}{\p u_{j}}} \Bigr) \\[4mm] \quad - {\rm C}_{j}^{i+j-l}
{\ds \frac{\p G}{\p u_{j}}} D^{i+j-l} \Bigl( {\ds \frac{\p F}{\p
u_{i}}}\Bigr) \Bigr], \qquad l=0, \dots, n+k-1,
\ea
\eeqn
 where $\ds {\rm C}_{q}^{p} = \frac{q!}{p! (q-p)!}$ and we assume that
$1/p!=0$ for negative integer $p$.

Let us also remind some facts concerning the formal series in
powers of $D$ (see e.g.  \cite{s}, \cite{mik} for more
information), i.e. the expressions of the form
 \beqn \label{fs}
 {\rm H}= \sum\limits_{j= -\infty}^{m} h_{j}
(x,t,u, u_{1}, \dots) D^{j}. \eeqn The greatest integer $m$ such
that $h_{m} \neq 0$ is called the degree of formal series ${\rm
H}$ and is denoted by $\deg {\rm H}$. For any formal series ${\rm
H}$ of degree $m$ there exists unique (up to the multiplication by
$m$-th root of unity) formal series ${\rm H}^{1/m}$ of degree $1$
such that $({\rm H}^{1/m})^{m} ={\rm H}$. Now we can define the
fractional powers of ${\rm H}$ as ${\rm H}^{l/m} = ({\rm
H}^{1/m})^{l}$ for all integer $l$. The key result here is
that\looseness=-2
 \beqn \label{cfs}
 \lbrack {\rm H}^{p/m}, {\rm H}^{q/m} \rbrack = 0
\eeqn
for all integer $p$ and $q$.

The formal symmetry of Eq.(\ref{eveq}) of rank $l$ is the formal
series ${\rm R}$, satisfying the relation \cite{s}
 \beqn \label{frmsym}
\deg  (\p {\rm R}/\p t + \nabla_{F}({\rm R}) - [F_{*}, {\rm R}])
\leq \deg F_{*} +\deg {\rm R} - l.
 \eeqn
The commutator of two formal symmetries of ranks $l$ and $m$
obviously is again a formal symmetry of rank not lower than
$\min(l,m)$, and thus the set $FS_r$ of all formal symmetries of
given equation (\ref{eveq}) of rank not lower than $r$ is a Lie
algebra. Like for the case of symmetries, we shall denote by
$FS_{r}^{(k)}$ the set of formal symmetries of degree not higher
than $k$ and of rank not lower than $r$. Note that if $G$ is
symmetry of order $k$, then by virtue of Eq.(\ref{cr3}) $G_{*}$ is
the formal symmetry of degree $k$ and rank $k + n -\deg \nabla_{G}
(F_{*})$. \looseness=-2

\section{Explicit form of symmetries and formal symmetries}
 In this section we shall consider a symmetry $G$ of order
$k \geq n_0$. The successive solving of Eq.(\ref{cr4}) for
$l=k+n-1,\dots, n_{0}+n-1$ yields Eq.(\ref{lead}) and
 \beqn \label{der2}
 \p G /\p u_i
= c_{i} (t) \Phi^{i/n} +\sum\limits_{p=i+1}^{k}
\sum\limits_{r=0}^{\lbrack \frac{p-i}{n-1} \rbrack } \chi _{i,p,r}
(x,t,u,u_1, \dots, u_{k}) \p^{r} c_{p} /\p t^{r}\!,
 \eeqn
 where $i=n_{0},\dots, k-1$ and $c_{i} (t)$ are some functions of $t$.
In particular, it may be easily shown that for $k > n+n_{0}-2$
\beqn \label{time}
\chi_{k-n+1,k,1} = (1/n) \Phi^{(k-n+1)/n}
D^{-1} (\Phi^{-1/n}).
\eeqn

Using Flach's theorem \cite{flach} and Eqs.(\ref{cr4}), (\ref{cfs}),
(\ref{time}), we have obtained \looseness=-1
\begin{theo}
For any symmetry $G$ of Eq.(\ref{eveq}) of order $k > n+n_{0}-2$
 \beqn \label{strg0a}
  \ba{l}
G_{*}= {\rm N} + \sum\limits_{j=\max(k-n+1,n_0)}^{k-1} d_{j} (t)
F_{*}^{j/n} + c_{k} (t) F_{*}^{k/n} \\[3mm] + \frac{k}{n^2} c_{k}
(t) D^{-1} (\Phi^{-1-1/n} \dot \Phi) F_{*}^{(k-n+1)/n}\\[3mm] +
(1/n) \dot c_{k} (t) D^{-1} (\Phi^{-1/n}) F_{*}^{(k-n+1)/n},
 \ea
 \eeqn
 where $d_{i}(t)$ are some functions of $t$ (in fact they are
 linear combinations of $c_{\max(k-n+1, n_0)}(t), \dots, c_{k}(t)$) and
${\rm N}$ is some formal series, $\deg {\rm N}< \max(k-n+1,n_0)$.
Likewise, for $n_0 \leq k \leq n+n_{0}-2$ Eq.(\ref{strg0a})
remains true, if two last terms on its right hand side are
rejected.
\end{theo}

Dot here and below stands for the partial derivative with respect
to $t$.

The analysis of Eq.(\ref{frmsym}), similar to the above analysis
of Eq.(\ref{cr3}), yields
\begin{theo}
For any formal symmetry ${\rm R}$ of (\ref{eveq}) of degree $k$
and of rank $r > n$\looseness=-2
 \beqn \label{strg0afsm}
  \ba{l}
{\rm R} = \tilde {\rm R} + \sum\limits_{j=k-n+1}^{k} d_{j} (t)
F_{*}^{j/n}+ \frac{k}{n^2} d_{k}(t) D^{-1}(\Phi^{-1-1/n} \dot
\Phi) F_{*}^{(k-n+1)/n}\\[3mm] +(1/n) \dot d_{k} (t) D^{-1}
(\Phi^{-1/n})F_{*}^{(k-n+1)/n},
 \ea
 \eeqn
where
$\tilde {\rm R}$ is
some formal series, $\deg \tilde {\rm R} < k-n+1$.
\looseness=-1
\end{theo}

\section{Structure of algebras of symmetries and formal symmetries}
Let us consider the Lie bracket $R = \{ P, Q \}$ of two symmetries
$P$ and $Q$. Obviously, $\ord R = \deg R_{*} \equiv r$, and we can
find from Eq.(\ref{lin}) $\p R/\p u_{r}$, which equals to the sum
of coefficients at $D^{r}$ on the right hand side of
Eq.(\ref{lin}).

The substitution of representations (\ref{strg0a}) for $P_{*}$ and
$Q_{*}$ into Eq.(\ref{lin}) yields after some computations the
following results:
\begin{theo} Let $P, Q \in S/S^{(n+n_{0}-2)}$, $\ord P = p$, $\ord Q
= q$. By virtue of Eq.(\ref{lead}) $\p P/\p u_{p} = c_{p}(t)
\Phi^{p/n}$, $\p Q/\p u_{q}= d_{q}(t) \Phi^{q/n}$.

Then $\ord \lbrace P, Q \rbrace \leq p+q-n$ and
 \beqn \label{crmain}
  \{P,Q\} = \frac{1}{n}\Phi^{\frac{p+q-n}{n}} u_{p+q-n} \Bigl(q \dot
c_{p}(t) d_{q}(t) - p c_{p}(t) \dot d_{q}(t)\Bigr)+\tilde R,
\eeqn
 where $\ord \tilde R < p+q-n$.
\end{theo}
\begin{cor} For all integer $p \geq n+n_{0}-1$ the spaces
$S^{(p)}$ are invariant under the adjoint action of $S^{(n)}$,
i.e. the Lie bracket of any symmetry from $S^{(p)}$ with any
symmetry from $S^{(n)}$ again belongs to $S^{(p)}$.
\end{cor}
\begin{theo} For all $p=0, \dots, n$ $S^{(p)}$ are Lie subalgebras in
$S$.
\end{theo}
%

Note that for $p=0,1$ the result of Theorem 4 is well known, while
for $p=2, \dots, n$ it is essentially new.

Theorem 3 shows that Lie algebra $S$ has Virasoro type structure
(if we forget about low order symmetries and consider just the
leading terms of symmetries). In particular,  one may easily
establish the existence of Virasoro (or hereditary) algebra
\cite{fu}, \cite{fu1} of time-independent symmetries and
mastersymmetries for many integrable equations (\ref{eveq}), using
Eq.(\ref{crmain}). More generally, Theorem 3 is very useful in the
proof of existence of infinite number of symmetries for given
evolution equation, starting from few initially found ones and
analyzing their commutators, as described in \cite{v}. Note that
for particular cases of KdV and Burgers equations it was proved in
\cite{v}.

Likewise, for formal symmetries we have proved the following
results:
\begin{theo}
 Let $\rm P, Q$ be formal symmetries of Eq.(\ref{eveq}),
$\deg {\rm P} = p$, $\deg {\rm Q} = q$, and the ranks of $P$ and
$Q$ are greater than $n$. By virtue of Theorem 2 ${\rm P}=
c_{p}(t)F_{*}^{p/n}+\tilde {\rm P}$ and ${\rm Q}= d_{q}(t)
F_{*}^{q/n} + \tilde {\rm Q}$, $\deg \tilde {\rm P}< p$, $\deg
\tilde {\rm Q} < q$.

Then $\deg \lbrack {\rm P, Q} \rbrack \leq p+q-n$ and
\beqn \label{crfsm}
  \lbrack {\rm P,Q} \rbrack = -\frac{1}{n}F_{*}^{\frac{p+q-n}{n}}\Bigl(q
  \dot c_{p}(t) d_{q}(t) - p c_{p}(t) \dot d_{q}(t)\Bigr)+\tilde {\rm R},
\eeqn
where $\deg \tilde {\rm R} < p+q-n$.
\end{theo}
\begin{cor} For all integer $p$ the spaces $FS_{r}^{(p)}$ are invariant
under the adjoint action of $FS_{r}^{(n)}$, provided $r > n$.
\end{cor}
\begin{cor} For all integer $p \leq n$, $r >n$ $FS_{r}^{(p)}$ are Lie
subalgebras in $FS_{r}$.\looseness=-1
\end{cor}

\section{Some possible generalizations}
 Our results may be easily generalized to the case of systems of
 evolution equations of the form (\ref{eveq}), when $u, u_{1}, u_{2},
 \dots, F, G, P, Q$ become $m$-component vectors, while $c_{p}(t), d_{q}
 (t)$ become
$m \times m$ matrices, which should commute with the $m \times m$
matrix $\Phi = \p F/\p u_{n}$, etc. (see \cite{s} for more
information). Namely, the second part of Theorem~1, concerning the
symmetries of order $k$, $n_{0} \leq k \leq n+n_{0}-2$, Theorem~3,
Theorem~4 for $n_{0} \leq 1$, Corollary~1, Corollary~3 for $p < n$
hold true, provided all the eigenvalues of $\Phi$ are distinct.
The part of Theorem 1, concerning the symmetries of order $k >
n+n_{0}-2$, Theorems~2,4,5, Corollaries~2 and 3 hold true if, in
addition to the above, $\det \Phi \neq 0$. For the case,when $\det
\Phi=0$, we have proved instead of the first part of Theorem~1 the
following result:\looseness=-1
\begin{theo}
If $\det \Phi =0$ and all the eigenvalues of $\Phi$ are distinct,
for any symmetry $G$ of system (\ref{eveq}) of order $k >
n+n_{0}-2$
\beqn
\label{strdeg} G_{*}= {\rm N} + \sum\limits_{j=k-n+2}^{k} d_{j}
(t) F_{*}^{j/n},
\eeqn
 where $d_{j}(t)$ are some matrices, commuting with
 $\Phi$, and ${\rm N}$ is some formal series (with matrix coefficients),
 $\deg {\rm N} < k-n+2$.
\end{theo}
In its turn, Theorem 2 is replaced by the following
\begin{theo}
If $\det \Phi =0$ and all the eigenvalues of $\Phi$ are distinct,
for any formal symmetry ${\rm R}$ of system (\ref{eveq}) of degree
$k \geq 0$ and of rank $r > n$\looseness=-2 \beqn \label{strdeg1}
{\rm R}= {\rm \tilde R} + \sum\limits_{j=\max(k-n+2,0)}^{k} d_{j}
(t) F_{*}^{j/n}, \eeqn
 where $d_{j}(t)$ are some matrices, commuting with
 $\Phi$, and ${\rm \tilde R}$ is some formal series (with matrix
coefficients),
 $\deg {\rm \tilde R} < \max(k-n+2,0)$.
\end{theo}

As a final remark, let us note that
Theorems 1 -- 7 and Corollaries 1 -- 3 may be also extended (under
some extra conditions) to the symmetries and formal symmetries,
involving nonlocal variables. We shall discuss this in more detail
in separate paper.

\section*{Acknowledgements}
I am sincerely grateful to Profs.~B.~Fuchssteiner and
V.~V.~Sokolov for sti\-mu\-lating discussions. \looseness=-1

\end{document}